\documentclass[a4paper]{article}
\usepackage[dvips]{graphics,epsfig}

\newcommand{\ybox}[2]	{
 \begin{center}
 \resizebox{!}{#1\textheight}
{\includegraphics{#2.eps}}
 \end{center}		}

\begin{document}
\thispagestyle{empty}

\title{Mass Composition of Cosmic Rays 
in the Range \\
$2 \times 10^{17} - 3 \times 10^{18}$ eV 
Measured with the Haverah Park Array}

\author{M. Ave$^1$, L. Caz\'on$^2$, J.A. Hinton$^1$\footnote{now at:
Max-Planck-Institut f\"ur Kernphysik, Heidelberg, Germany},
J. Knapp$^1$, J. Lloyd-Evans$^1$, and 
A.A. Watson$^1$ \\[2mm]
\normalsize
$^1$ Dept. of Physics and Astronomy, University of Leeds, Leeds LS2 9JT, UK\\
\normalsize
$^2$ Dept. de F\'\i sica de Part\'\i culas, Universidad de Santiago,
15706 Santiago de Compostela, Spain}

\date{}
\maketitle

\begin{abstract}
At the Haverah Park Array a number of air shower observables were
measured that are relevant to the determination of the mass composition
of cosmic rays. In this paper we discuss measurements of the risetime of
signals in large area water-Cherenkov detectors and of the lateral
distribution function of the water-Cherenkov signal.  The former are
used to demonstrate that the CORSIKA code, using the QGSJET98 model,
gives an adequate description of the data with a low sensitivity, in
this energy range, to assumptions about primary mass.  By contrast the
lateral distribution is sufficiently well measured that there is mass
sensitivity. We argue that in the range 0.2--1.0 EeV the data are well
represented with a bi-modal composition of $(34\pm2)$\% protons and the
rest iron. We also discuss the systematic errors induced by the choice
of hadronic model.
\end{abstract}

\section{Introduction}

High-energy cosmic rays are measured via extensive air showers of
secondary particles they produce in the Earth's atmosphere. Specific
observables depend in a complex way on primary mass and energy, which
must be understood to interpret air shower data. Some measurable
parameters at ground can be used to obtain the primary energy of the
incoming cosmic rays without large systematic uncertainties due to the
unknown composition \cite{spectrum}.  The main effect of a variable mass
on showers with the same initial energy, $E$, is a characteristic shift
of the position of the maximum of shower development in the atmosphere,
$X_{\rm m}$.  Cosmic ray nuclei share their energy amongst $A$ nucleons
and, therefore, their showers can, to a first approximation, be
described as a superposition of $A$ nucleon-induced subshowers with an
energy of $E/A$ each.  Thus, showers of heavy primaries are less
penetrating, and tend to develop higher up in the atmosphere, than
nucleon showers of the same energy.  Therefore, $X_{\rm m}$ is an
observable with a strong connection to the primary mass.  However,
additionally, the depth of maximum also depends on energy.  The higher
the energy the longer the shower and the deeper is $X_{\rm m}$. This is
described by the elongation rate, d$X_{\rm m}$/d$\log E$, which is
usually quoted as the shift in $X_{\rm m}$ per change of energy by a
factor 10.  In practice, apart from the unique case of fluorescence
detectors, there is no way to measure $X_{\rm m}$ directly.  Mostly
other quantities are recorded that can be shown to have a correlation to
the height of the shower development.

Here we report an analysis of mass composition in the energy range
0.2--1.0 EeV that has been performed with data from the Haverah Park
extensive air shower array. The Haverah Park array was a 12 km$^2$ air
shower array consisting of water tanks that acted as Cherenkov
detectors. It was operational from 1967 to 1987.

At Haverah Park a number of observables were measured which are relevant
to the determination of the mass composition of cosmic rays above
$2\times 10^{17}$~eV. In particular, the two parameters, $\eta$ and
t$_{1/2}$, which are sensitive to the longitudinal development of
showers, have been studied in detail. $\eta$ describes the steepness of
the lateral distribution function of the signal observed in the
water-Cherenkov detectors. $\eta$ was measured with high precision using
a portion of the array with a much denser detector arrangement, the
so-called infill array \cite{coy}.  For 1425 events in the energy range
$2\times10^{17}$ to $3\times10^{18}$~eV information from infill
detectors was available and $\eta$ could be determined. These were
recorded between 1977 and 1981.

\begin{figure}
\ybox{0.35}{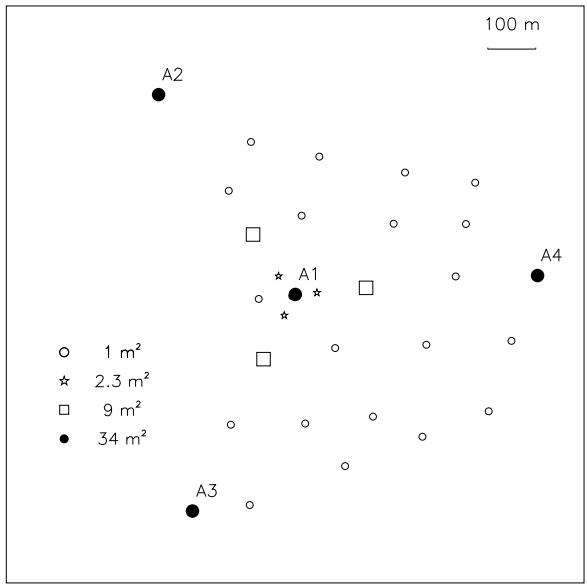}
\caption{\small The inner part of the Haverah Park array, the so called infill array.}
\label{infill}
\end{figure}

The second mass-sensitive parameter, the risetime t$_{1/2}$, is a
measure characterising the spread of the arrival times of individual
particles at a given detector. It is evaluated from the integrated
signal of all particles belonging to a shower and defines the time
interval in which the integrated signal rises from 10\% to 50\%.
Risetimes can be sensibly determined only in large-area
detectors. Therefore they are obtained from the four 34~m$^2$ detectors,
A1--A4 (Fig. \ref{infill}). Risetime data were obtained from $>7000$
events and a total of 13000 detector signals at core distances of more
than 300~m. The events used here span a range of energies from
$2\times10^{17}$ to $2\times 10^{19}$~eV.
 
The analysis of risetimes of events with known core position, arrival
direction and energy established the first evidence of shower-to-shower
fluctuations at high energies \cite{watson74}.  This was supported by
the demonstration of a correlation of t$_{1/2}$ with the steepness of
the lateral distribution,
which could be understood as a consequence of both variables being
dependent on $X_{\rm m}$. In the 1970s and 1980s it was not possible to
make accurate predictions of the t$_{1/2}$ expected for different
primary masses because the problem could not be solved analytically and
a full 4-dimensional numerical simulation was beyond the capabilities of
the computing power accessible at that time.

\begin{figure}
\ybox{0.35}{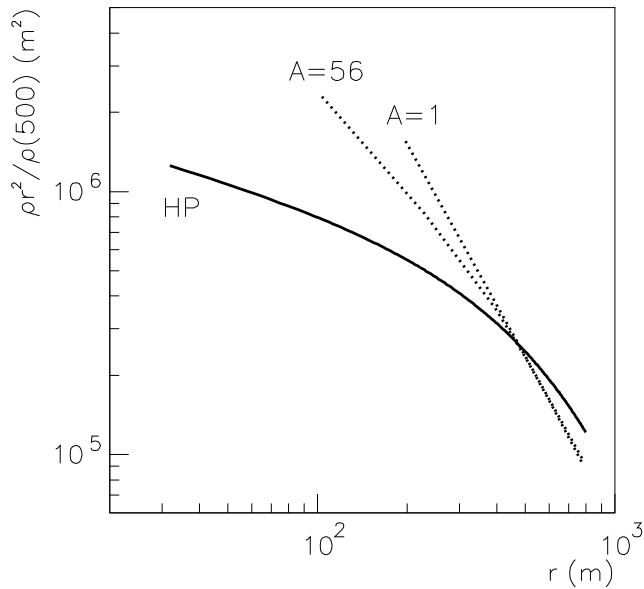}
\caption{\small A comparison of Haverah Park data and the calculations for p
and Fe induced showers from \cite{gaisser}. $\rho$ is the
water-Cherenkov signal.}
\label{gaisser}
\end{figure}

The analysis of $\eta$ data also showed evidence of fluctuations very
much larger than could be attributed to the experimental uncertainties
\cite{paris}.  The dependence of $\eta$ on the depth of the shower
development in the atmosphere could not be established directly from the
data.  Furthermore, a comparison of the measured average value of $\eta$
with a highly regarded model calculation of the time \cite{gaisser}
showed strikingly poor agreement: a mean cosmic-ray mass very much
heavier than iron was required to fit the data (see Fig. \ref{gaisser}).

The situation is now very different and
a model can be found that describes the data rather well.  To interpret
the data detailed simulations with the CORSIKA and GEANT programs have
been performed.  CORSIKA \cite{corsika} is a modern air shower model
that simulates particle production and propagation in the atmosphere
fully. It employs various models of high-energy hadronic interactions,
of which the most tested is QGSJET98 (Quark-Gluon-String model with
jets) \cite{qgsjet}.  The GEANT program \cite{geant} is used to account
for the detector response to particles of different type, energy and
impact angle.

\section{The Haverah Park Array} 

The Haverah Park array has been described in detail elsewhere
\cite{haverah}. The central part of the array is composed of four 34
m$^2$ water-Cherenkov detectors, A1--A4, spaced 500 m apart.  Six groups
of satellite arrays each of 4 $\times$ 14 m$^2$ lie near to a
circumference of radius 2 km around the centre so that density
information is available from all sites for showers that trigger the 500
m array. The 500 m array is sensitive to showers with primary energies
above $\approx 6\times10^{16}$ eV.  Another set of three detectors (9
m$^2$ each) surrounding the central detector (A1) form a 150 m array.
The so-called infill array consists of 23 small-area water-Cherenkov
detectors located in the central region of the 500 m array with an
irregular spacing of $\approx 150$ m. For the arrangement of detectors,
see Fig. \ref{infill}. Further small-area detectors were placed right
next to the 9 m$^2$ detectors and A1 for calibration purposes.

There were two kind of detectors: (i) 2.29 m$^2$ area galvanized
rectangular iron tanks with aluminum lids, identical to those used in
the 4 $\times$ 34 m$^2$ detectors: these detectors were placed inside
wooden huts with light roofs, and (ii) 1 m$^2$ area water tanks, made of
light-weight polyurethane foam and covered with fibre glass: these tanks
of hexagonal cross section, stood outdoors.  All tanks were filled with
water to a depth of 1.2 m and viewed by one photomultiplier with 100
cm$^2$ photocathode.  Detector areas larger than 2.29 m$^2$ were
achieved by grouping several of the 2.29 m$^2$ detectors
together. Sixteen of the modules form each of the detectors A1--A4.  The
signals from 15 of the 16 tanks were summed to provide the signal used
for triggering and for the density estimate. The 16$^{\rm th}$ tank in
each group was used to provide a low gain signal.  4 modules make up the
9 m$^2$ detectors and 6 modules form the 13 m$^2$ detectors.

The densities of Cherenkov photons per unit detector area (Cherenkov
densities) were recorded in terms of the average signal from a vertical
muon (1 vertical equivalent muon = 1 vem) per square metre.  This signal
corresponds to approximately 14 photoelectrons (pe) for Haverah Park
tanks. A trigger for all detectors is generated when the central
detector (A1) and at least two out of the three other A-sites (A2, A3,
A4) detect a particle density of $> 0.3$ vem/m$^2$. However, infill
signals had to be above 7 vem/m$^{2}$ to be recorded. Trigger rates were
monitored daily over the life of the experiment. After correction for
atmospheric pressure variations, the trigger rates were found to be
stable to better than 5\%.

A feature of the Haverah Park array, which is particularly important for
the shower front measurements, is the area of the detectors
A1--A4. These detectors allow large samples of the shower front to be
taken at widely separated points, thus minimizing the chance of
observing effects due only to local density fluctuations.  The temporal
response of the recording system to a step function was determined by
measuring the response to small showers that fell nearby and thus
produced short risetimes. The risetime of bandwidth limited pulses was
45 ns \cite{walker2}.

Cross-calibration of the detectors of the infilled array with the other
detectors could have been achieved using vertical muons, as for the
large-area detectors, so that they yield the same response to a purely
muonic signal.  However with such a cross-calibration one would have had
to correct for the different response of the detectors to the soft
component and for differences in optical and geometrical properties of
the detectors and so an alternative approach was adopted.  Signals from
individual 2.29 m$^2$ and 1 m$^2$ detectors, next to A1, were compared
with corresponding densities from the 34 m$^2$ detector, and an
appropriate conversion factor was obtained.  Similar calibration
procedures were used on simulated data for consistency.

\section{Model calculations}
\label{mc}

At present the CORSIKA program is effectively the standard tool for air
shower simulations, and it is successfully used by a variety of
different experiments over the energy range from 1--10$^{12}$
GeV. CORSIKA uses EGS4 (Electron Gamma Shower code) \cite{egs} for the
simulation of the electromagnetic interactions, and features a detailed
simulation of particle propagation and decay.  However, most important
for the development of air showers in the atmosphere are soft hadronic
interactions.  At present the only available theoretical approach to
model these is Gribov-Regge theory of multi-Pomeron exchange.  Several
interaction models based on this theory are available in CORSIKA, the
most successful being QGSJET \cite{qgsjet}. The QGSJET model also
contains the treatment of hard collisions and the production of
mini-jets, which become dominant at very high energy collisions, and
includes a realistic simulations of nucleus-nucleus collisions.  Its
parameters were tuned to fit a wide range of accelerator results and to
provide a secure extrapolation to the highest energies.  The combination
of CORSIKA with QGSJET seems to be able to describe experimental cosmic
ray results from Cherenkov telescopes at $10^{12}$ eV \cite{qgsjet_tev},
over measurements of the hadronic, muonic and electromagnetic shower
components in the knee region \cite{modtest}, to lateral distributions
and arrival times as will be shown in this work, up to air showers at
$10^{20}$ eV \cite{qgsjet_uhecr}.

Using the CORSIKA code with QGSJET98 we have generated a library of
proton and iron showers with zenith angles: $0^\circ$, $15^\circ$,
$26^\circ$, $40^\circ$ and $45^\circ$ at a primary energy of
$4~\times~10^{17}$ eV, and for a zenith angle of $26^\circ$ at energies
of 0.2, 0.4, 0.8, 1.6, and 3.2 EeV.  For each set 100 showers with the
same parameters have been simulated. Additionally we have generated sets
of 100 showers with oxygen and helium primaries at zenith angle
26$^\circ$ and energy 0.4 EeV.  In total 3600 showers were produced.
Statistical thinning of shower particles was applied at the level of
10$^{-6} \times$ E with a maximum particle weight limit of 10$^{-13}
\times$ E/eV.

Predictions of the risetimes were deduced from the time information of
each individual particle contained in the simulated shower library. The
time distribution of the signals produced by electromagnetic particles
for different core distances ($r$) was obtained for each simulated
shower.  The integral over the distribution was normalized to the
expected signal from electromagnetic particles at the given distance for
a 34 m$^2$ detector.  The arrival time distribution of muons has been
obtained for different $r$.  The expected number of muons at a given $r$
for a 34 m$^2$ detector has been computed and Poisson fluctuations
added.  The muons are sampled from the arrival time distribution to
obtain the time distribution of the muon signal in the detector. The
time distributions of the muon signal and the soft component were then
added and convoluted with the known system response to an instantaneous
pulse.  The risetime is then inferred from the result of the
convolution.  This procedure was repeated 100 times for each shower to
get the mean risetime as a function of $r$ and the expected spread of
the risetimes arising from Poisson fluctuations in the number of muons.
To increase the statistics further each CORSIKA shower was thrown 100
times onto the array with random core positions ranging out to 300 m
from the centre of the array. The risetime at each of the four 34 m$^2$
detectors is calculated from the parameterisations obtained above.  The
risetime at each detector is smeared according to an error that contains
contributions from Poisson fluctuations in the number of muons and from
the experimental measurement error ($\approx$ 3 ns). The core distance
of each detector was varied according to the error in the core position
described in \cite{spectrum}.  To reproduce the multiplicity of measured
risetimes per event in the data, we keep only the core position and the
risetime of some of the detectors.  We repeated this procedure for each
of the 100 CORSIKA showers in a set and the mean risetime versus $r$ is
obtained (Fig. \ref{risevsr}).

To obtain a value of $\eta$ for each simulated shower we have
convoluted the CORSIKA output with the detector response and fitted the
resulting lateral distribution function for muons and electromagnetic
particles.  Again, each shower was used 100 times with core positions
randomly scattered over the infilled area, obtaining the densities at
each detector with the parameterisations described above.  The densities
are modified according to Poisson fluctuations in the number of
particles. We tested whether an event meets the array trigger conditions
and, if so, the densities were fluctuated according to measurement
errors and recorded in the same format as real data.  The calibration of
the infilled array is reproduced with simulations to obtain the
conversion factor described in the previous section.  To obtain the
energy of the showers, either simulated or real, we need its relation to
the water-Cherenkov signal density at a core distance of 600 m,
$\rho(600)$, and the attenuation length $\lambda$. The procedure to
calculate these is described in ref. \cite{spectrum}. Here we have
followed the same procedure. The uncertainty in the core position, for
the $\eta$ analysis, is $\approx 5$ m for all energies and primary
masses tested. The energy resolution for proton primaries falls from
15\% at 0.4 EeV to 10\% at 6.4 EeV, while for iron primaries it is 10\%
and 7\%, respectively. These uncertainties include physical fluctuations
in $\rho(600)$ and measurement errors.

\section{Risetime analysis} 

\begin{figure}[b]
\ybox{0.30}{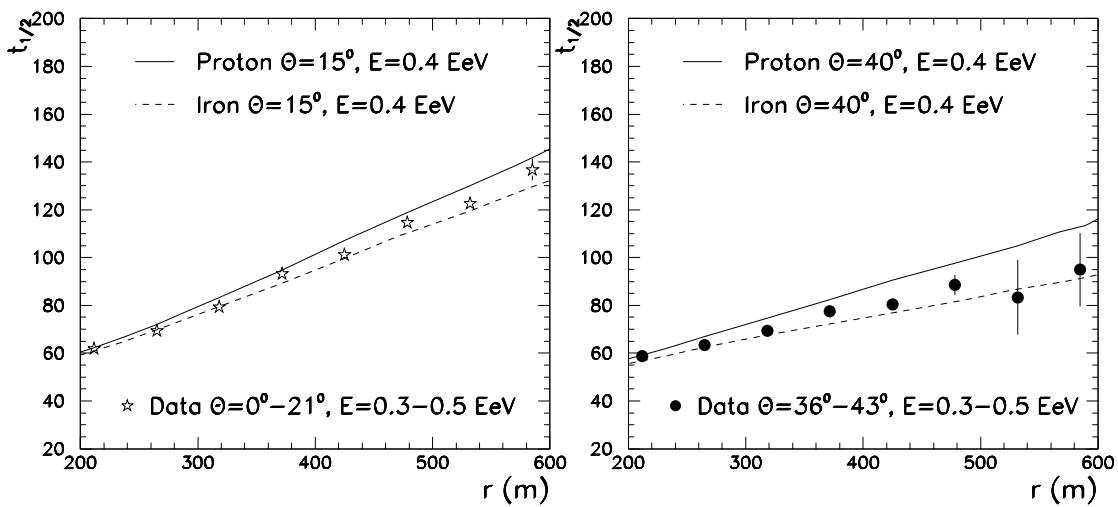}
\caption{\small Risetime versus distance to shower core as obtained in data and
in simulations for proton and iron primaries.}
\label{risevsr}
\end{figure}

The dependence of the risetime of the signal in a detector on shower
development arises largely because of geometry.  Particle scattering,
velocity differences and geomagnetic deflections are second order
effects: the geometrical sensitivity is enhanced at large distances. The
set of nuclear interactions that produce particles observed at an
off-axis detector may be regarded as lying on a line source.  There are
two important properties of this line source that relate to mass
composition and affect the observed risetime. These are its position in
the atmosphere and its length.

Previous studies of t$_{1/2}$ have revealed its dependence on $\theta$,
energy and core distance, and used deviations from mean values as
measures of the variations of $X_{\rm m}$ with energy to get results on
mass composition. The variations deduced \cite{walker} were in good
accord with measurements made subsequently and directly by the Fly's Eye
group \cite{cassiday}.

The strategy for the present analysis is different: the risetime as a
function of core distance for different bins of $\theta$ and $E$ are
compared directly with predictions from Monte Carlo simulations for
different primary masses.  In the energy and distance ranges
($250<r<500$ m) of interest here, where $\eta$ has been obtained with
high accuracy, the sensitivity of the risetime technique for the
extraction of mass information is rather limited.  The technique is,
however, expected to be very effective for mass separation at greater
distances and in larger showers where the mean risetime difference
predicted between proton and iron showers is large and easy to
measure. The results of such a study will be described elsewhere. At
small distances limitations of bandwidth mean that relatively large
samples of showers were required to establish the dependence of risetime
on energy and zenith angle and to demonstrate shower-to-shower
fluctuations \cite{walker,watson74}.  However this relative
insensitivity to mass can be turned to advantage by using the data to
test whether a shower model is able to predict average risetimes
accurately over the distance range discussed here.  In
Fig. \ref{risevsr} we compare the mean risetime with distance, for two
zenith angle bands, with model predictions.  The data lie between the
predictions for protons and iron: it is not wise to deduce any mass
information from these plots because of known systematic uncertainties
in the data and the models at the few nanosecond level. The
flattening-off evident at the smaller core distances is a consequence of
the low bandwidth of the 1970s recording system.

Two cuts were applied to the data and to the simulation results to
obtain Fig. \ref{risevsr}: the density in a particular detector had to
be greater that 1 vem/m$^{2}$ (to reduce sampling fluctuations), and the
position of the core had to be closer than 300 m from the central
triggering detector (to avoid large core errors).

We infer from this analysis that the CORSIKA code, using QGSJET98
physics, gives a very adequate description of the mean shower risetime
as a function of distance from 200--800 m and over the zenith angle
range 15$^\circ$--40$^\circ$.  This is the first time that such
agreement has been demonstrated.  While the comparisons of
Fig. \ref{risevsr} do not provide proof that the QGSJET98 model is
correct they do give us reasonable confidence in using this model to
attempt to interpret other data that are expected to show mass
sensitivity.

\section{Lateral distribution analysis}

\subsection{Event reconstruction}

 First the shower direction was determined using the arrival time
information of the four central triggering detectors. The particle
density information was then analysed to find the shower core.  The
lateral distribution, parameterised with respect to the known core
position and the primary energy, is estimated from the particle
densities and the form of the lateral distribution using the correlation
between $\rho(600)$ and energy.  At a given energy the lateral distribution was found to
be well described experimentally by the modified power law:
\begin{equation}
\rho(r)=k~r^{-(\eta+r/4000~{\rm m})} \\
\label{ldf}
\end{equation}
and fits well to the average data in the distance range 80--800 m.
$\eta$ was shown to vary with zenith angle as $\eta=3.78
-1.44~\sec\theta$.

Due to the small number of densities usually available without the
infill array, it was not possible to fit reliably values of $\eta$ for
each individual shower.  Therefore, the dependence of $\eta$ on zenith
angle was obtained by calculating the average water-Cherenkov lateral
distribution function for different zenith angle bins.  However, the
great number of densities available with the infill array allows study
of the fluctuations of the lateral distribution on a shower-by-shower
basis.

The original algorithm used to determine the shower parameters is
described as follows. For a given lateral distribution function 
geometrical reconstruction
provides a unique core location using only a small number of density
measurements.  For each selected event the core was found using 3 or 4
detectors surrounding the probable core position. The geometrical mean
distance of the shower core from these ``ringing'' detectors was in the
range $50 < r_{\rm n} < 200$ m. Only densities at distances $>r_{\rm n}$
were used to determine the lateral distribution function using a $\chi^2$ minimisation
technique. An iterative procedure was used to find the best value of
$\eta$ for each event using as a starting point the energy independent
approximation of $\eta$ given above.  A multiple linear regression was
used to determine the dependence of $\eta$ \cite{england} on zenith
angle and energy. The result was
$$
\eta = a - b\, (\sec\theta-1)+ c \log(E/10^{17}{\rm eV})
$$
with $a = 2.20\pm0.01$, $b = 1.29\pm 0.05$, and $c = 0.165\pm0.022$.
The difficulty of the ringing analysis is the implementation of the
algorithm. The optimum set of ringing detectors was chosen
subjectively. Several factors were considered in the choice of ringing
analysis but the experience of the observer was also important. For the
purposes of this analysis, in which a large statistical sample of
simulated events is generated to compare with data, another technique
described below was adopted to find the optimum shower parameters, so
that the role of the observer was removed.

\begin{figure}
\ybox{0.35}{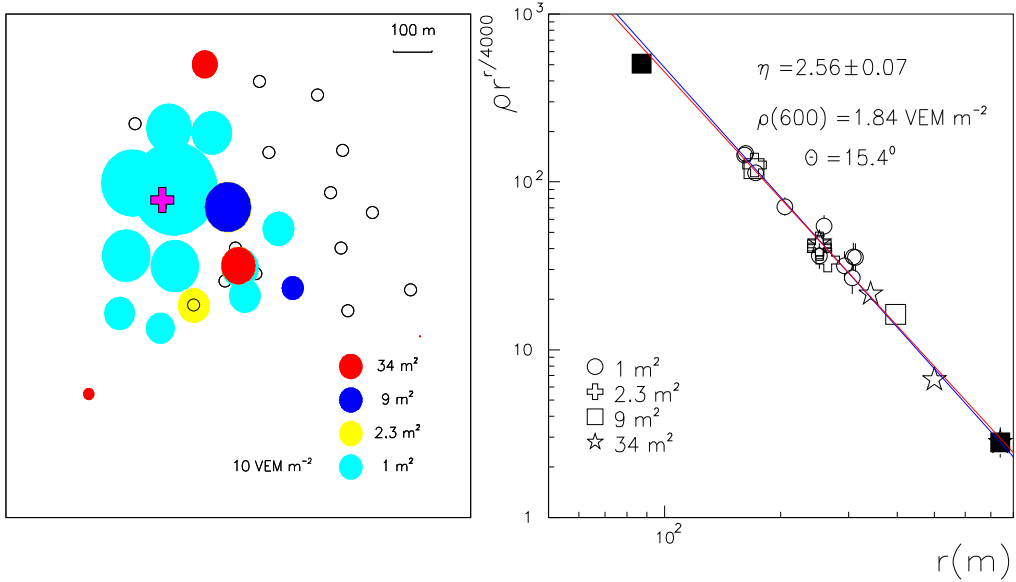}
\caption{\small Example of the reconstruction of an infill event. Left panel:
Projection of the array into the shower plane with recorded densities
shown as circles with radius proportional to the logarithm of the
density. The detector areas are indicated by grey scales. The detectors
are displayed in the plane perpendicular to the shower axis. Right
panel: Fitted lateral distribution function. The two lines correspond to
the lateral distribution function 
as obtained with ringing analysis and with the method described
in the text. The difference in the value of $\eta$ for this particular
event is 0.04. The abscissa shows a quantity that linearises the lateral
distribution function
given in Eq. \ref{ldf}. Filled symbols indicate detectors with signals
above saturation or below threshold.}
\label{event}
\end{figure}

The algorithm used here is a grid search over many different core
position. For a given core position the best value of $\eta$ and $k$ are
found through an analytical fit and the $\chi^2$ function is
computed. We used only detector densities above threshold and below
saturation in the range 80--800 m. Densities above saturation and below
threshold do not improve the fit because of the large number of recorded
densities available in this data set. Fig. \ref{event} shows an example
of a reconstructed event using this method and the ringing analysis. The
difference in the value of $\eta$ between the two analyses were found to
be $\approx 0.08$, which agrees with the measurement error
reconstruction in $\eta$ for the two techniques ($\approx 0.08$).

The selection criteria applied to the data are the following:
\begin{itemize}
\item Zenith angles in the range 0$^\circ$--45$^\circ$.
\item In every shower the core must be located within the infilled area.
\item Showers in which the largest density is at one of the boundary detectors
are excluded.
\item The value of $\chi^2$ from the $\eta$ fitting procedure should
correspond to the ``goodness of the fit'' having a probability greater
than 1\%.
\end{itemize}
After the application of these criteria 1351 showers remain from the
initial number of 1450. This is the data set we use to compare with
model calculations.

\subsection{Model comparison}

We use two independent methods to extract information about mass
composition from the lateral distribution function. The first one is
based on the $\chi^2$ method and the second one on the likelihood
approach. The first technique is more direct and allows a visual check
of the procedure, but requires the use of zenith angle cuts in our data
set, reducing thus the statistics.  The second method allows us to
combine all the zenith angles to predict the mass composition in
different energy ranges. We will show that the two methods are
consistent but we will give our final results in terms of the second
method.

\subsubsection{Method 1}
\label{method1}

\begin{figure}[b]
\ybox{0.3}{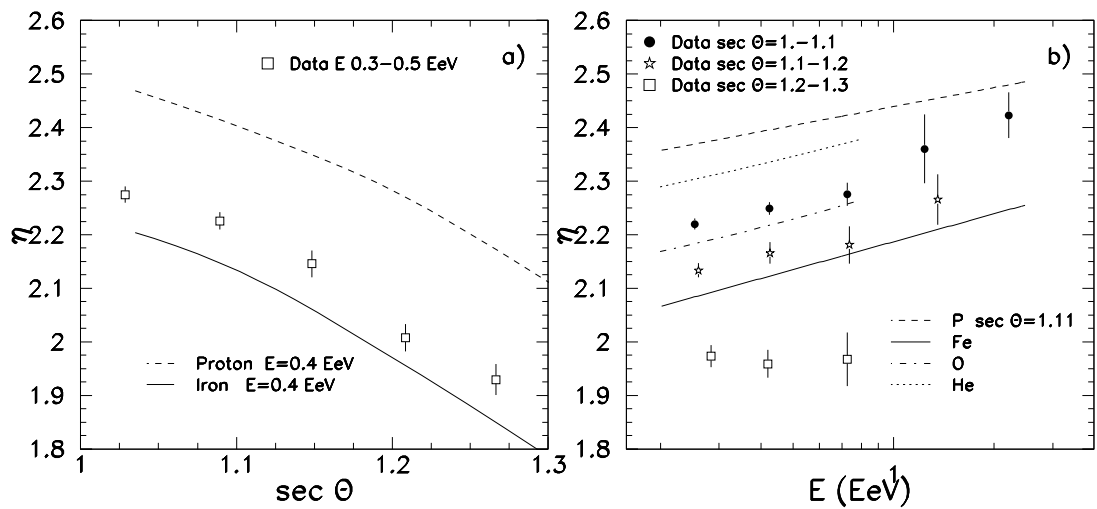}
\caption{\small Right panel: evolution of $\eta$ with zenith angle; the
simulation results correspond to a zenith angle of 26$^\circ$, the data
is binned in zenith angle bands. Left panel: evolution of $\eta$ with
zenith angle for data and simulations.}
\label{etaen}
\end{figure}

The simulated events generated from CORSIKA showers are analysed with
the same algorithms as real data and the same cuts are applied. In
Fig. \ref{etaen}(a) the variation of $\eta$ with zenith angle is shown
for events with energies in the range 0.3--0.5 EeV.  It is evident that,
if the mean mass lies between the limits of proton and iron, the lower
energy data are well described by the QGSJET98 model.

\begin{figure}
\ybox{0.35}{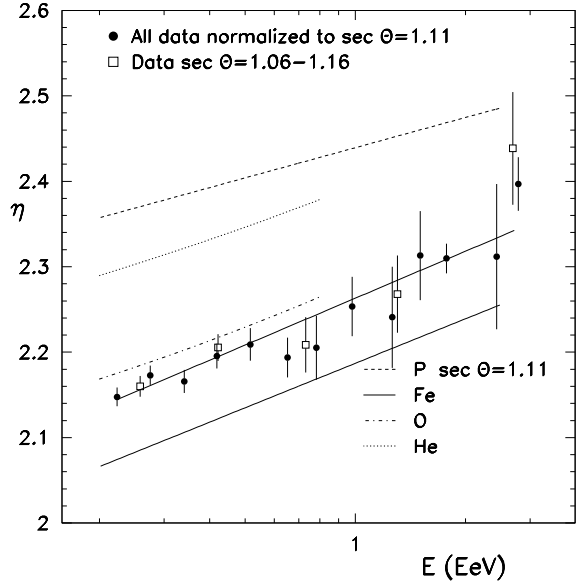}
\caption{\small Evolution of $\eta$ with energy for data and simulation
results.}
\label{etazen}
\end{figure}

In Fig.  \ref{etaen}(b) and \ref{etazen} we show the variation of $\eta$
with energy.  In Fig. \ref{etaen}(b) data in three zenith angle ranges
are displayed. The model calculations shown for the four mass species
are for the boundary between the two most vertical zenith angle bands.
In Fig. \ref{etazen} the data have been normalised to $26^\circ$ for
comparison with calculations at the same angle.  The agreement between
the normalised data and a smaller data set in the range $1.06 <
\sec\theta < 1.16$ is good and gives confidence in the normalisation
procedure.  A linear fit to the normalised data is shown: the reduced
$\chi^2$ is 1.1.  The data might be described with a single mass
component, independent of energy, or by a mixture of several mass
components.  To distinguish between these possibilities we use data on
the spread of $\eta$.
\begin{figure}
\ybox{0.5}{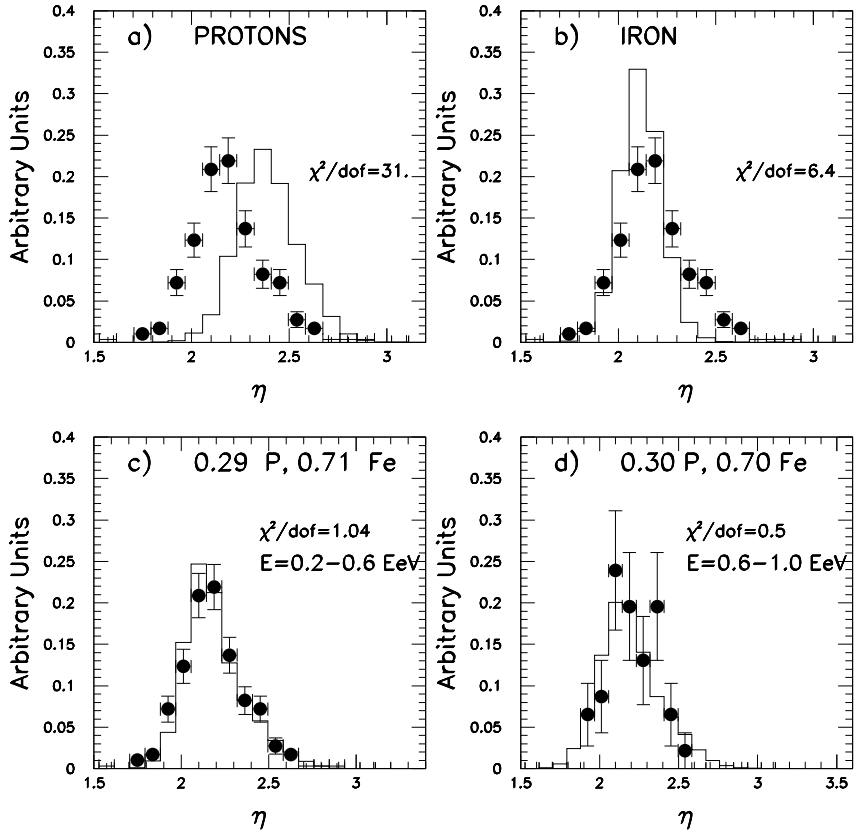}
\caption{\small Distributions of $\eta$ for experimental data, and model
predictions with different primary masses.}
\label{etadist}
\end{figure}

Although the average properties of showers do not vary strongly with
primary mass, fluctuations in the longitudinal development are
substantially larger for proton than for iron-initiated showers. The
smaller fluctuations for nucleus-induced showers is a direct consequence
of the superposition of A nucleonic showers, produced after the initial
nucleus fragments, which tends to average out extreme fluctuations that
might occur in individual subshowers.  To study fluctuations in $\eta$,
two $\eta$ distribution were constructed: (i) showers with 1.06$< \sec
\theta < 1.16$ and 0.2 EeV$ < E < 0.6$ EeV (292 events), and (ii)
showers with $1.04 < \sec \theta < 1.18$ and 0.6 EeV$ < E < 1.0$ EeV (46
events).

To compare with simulations, we account for all sources that could
contribute to the spread of $\eta$, namely fluctuations in the
longitudinal development of the shower and reconstruction errors. These
are automatically taken into account, since fluctuations are contained
in the simulated event sample, and the same reconstruction methods are
used for simulated events as for real ones. The typically root mean
square spread from ``shower-to-shower fluctuations'' is 0.12 for proton
showers and 0.05 for iron showers, and from the measurement error in
$\eta$ is $\approx$ 0.08.

In Fig. \ref{etadist}(a) and (b) the experimental data in the energy
band 0.2--0.6 EeV are compared with predictions for proton and iron
beams. The reduced $\chi^2$ for the fits are 31 and 6.4,
respectively. The tail at large $\eta$ in the comparison with iron
indicates that some light nuclei are required to fit the measurements.
For helium and oxygen the corresponding $\chi^2$ values are 20.0 and
7.0.  A fit to a dual component mixture is shown in
Fig. \ref{etadist}(c).  A mixture with ($29\pm 4$)\% of protons and a
corresponding amount of iron is a very good fit ($\chi^2/{\rm dof} =
1.04$): the addition of small amounts of helium and oxygen ($< 2$\%)
gives $\chi^2/{\rm dof} = 0.8$.  In Fig. \ref{etadist}(d) we show 46
events in the energy range 0.6--1.0 EeV and zenith angle range $1.04 <
\sec\theta < 1.18$: a two component fit gives ($30\pm 10$)\% protons.

We thus conclude that there is no evidence for any change of mass with energy
in the range studied and that Fe is the dominant component.  The data of
Fig. \ref{etazen} do not exclude the possibility of the Fe component getting
larger as the energy reaches 1 EeV and then smaller beyond.

\subsubsection{Method 2}

The zenith angle cuts used in the procedure described above do not allow
us to use all the data available in a given energy range to extract mass
composition. For this reason we have developed an additional technique,
using the likelihood method, that exploits the data more effectively.

For each set of CORSIKA showers at a given energy and zenith angle we have
generated an $\eta$ distribution. Each set consists of 100 CORSIKA showers,
and each shower was thrown randomly onto the array. We have reconstructed it
with the same procedure as for the data, obtaining an $\eta$ distribution that
includes the effect of shower-to-shower fluctuations and experimental
reconstruction. We have fitted each resulting $\eta$ distribution to a
Gaussian obtaining the mean $\mu_\eta$ and width $\sigma_\eta$.
 
\begin{figure}
\ybox{0.3}{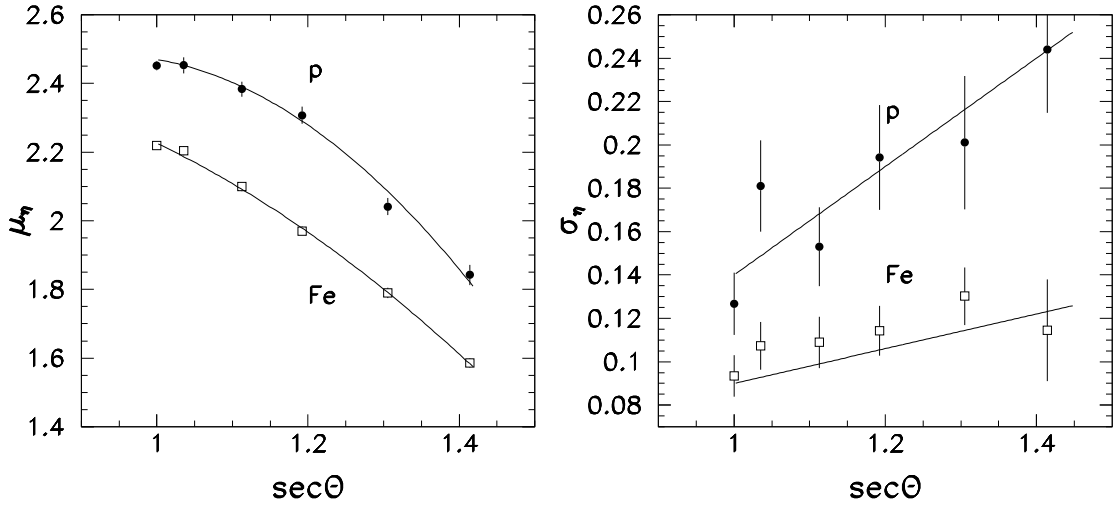}
\caption{\small Variation of $\mu_\eta$ and $\sigma_\eta$ with zenith angle for
proton and iron primaries at an energy of 0.4 EeV. The lines have the
form of Eqs. \ref{mean}--\ref{sigma2}, and are fitted to the results
obtained from simulations. The top lines correspond to proton primaries
and the bottom lines to iron primaries.}
\label{zenfit}
\end{figure}

\begin{figure}
\ybox{0.3}{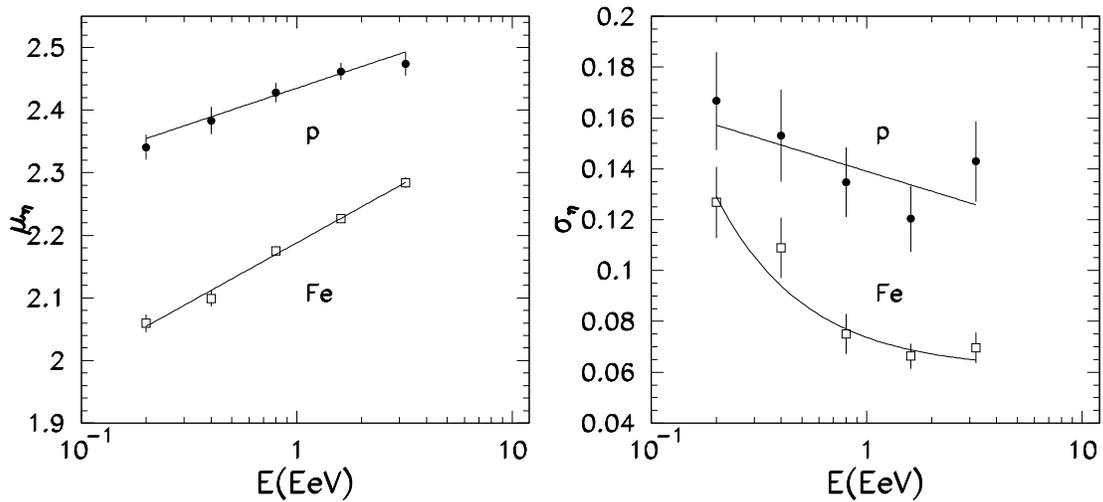}
\caption{\small Variation of $\mu_\eta$ and $\sigma_\eta$ with primary energy
for proton and iron primaries at a zenith angle of 26$^\circ$. The lines
have the form of Eqs. \ref{mean}--\ref{sigma2}, and are fitted to the
results obtained from simulations.}
\label{enfit}
\end{figure}

We can parameterise the values of $\mu_\eta$ and $\sigma_\eta$ as a function
 of zenith angle at a fixed energy (0.4 EeV), and as a function of the energy
 at a fixed zenith angle (26$^\circ$), for proton and iron
 primaries. Fig. \ref{zenfit} and \ref{enfit} show these fits, which are very
 satisfactory. Therefore we can predict the values of these parameters for any
 energy and zenith angle.

The values of $\mu_\eta$ and $\sigma_\eta$ are given by:
\begin{eqnarray}
\mu_\eta(\theta,E)&=&a_1+a_2~\Delta_\theta +a_3~\Delta_\theta^2+a_4~\log_{10}(E/0.4 {\rm EeV}) \label{mean} \\ 
\sigma_\eta(\theta,E, p)&=&b_1+b_2~\Delta_\theta + b_3~\log_{10}(E/0.4 {\rm EeV})\label{sigma} \\
\sigma_\eta(\theta,E,Fe)&=&b_1+b_2~\Delta_\theta + b_3~\exp(b_4~\log_{10}(E/0.4 {\rm EeV})) \label{sigma2}
\end{eqnarray}
with 
$\Delta_\theta = \sec\theta-\sec 26^\circ$,
and the constants obtained from the fits shown in Figs. \ref{zenfit} and \ref{enfit}
for proton and iron primaries are given in Table \ref{tablemu}. It
should be noticed that we use a different functional form for $\sigma_\eta$
depending on the primary. $\sigma_\eta$ is decreasing with increasing energy
as the shower-to-shower fluctuations are reduced, but it cannot have a value
smaller than the reconstruction error of $\eta$ ($\approx 0.08$), which is
constant with energy. This is the reason for the flattening of the value of
$\sigma_\eta$ for iron primaries at high energies in Fig. \ref{enfit}.

\begin{table}
\begin{center}
\begin{tabular}{|c|c|c|c|c|}
\hline
$\mu_\eta$ & \mbox{$a_1$} & \mbox{$a_2$}& \mbox{$a_3$} & \mbox{$a_4$} \\
\hline
p  & 2.38 $\pm$ 0.01   & -1.00 $\pm$ 0.10 & -2.8 $\pm$ 0.6 & 0.12 $\pm$ 0.02  \\  
\hline
Fe & 2.10 $\pm$ 0.01   & -1.33 $\pm$ 0.06 & -1.2 $\pm$ 0.3 & 0.19 $\pm$ 0.01  \\  
\hline
\hline
$\sigma_\eta$ & \mbox{$b_1$} & \mbox{$b_2$}& \mbox{$b_3$} & \mbox{$b_4$} \\
\hline
p  & 0.16$\pm$ 0.01   & 0.25 $\pm$ 0.07 & -0.03 $\pm$ 0.02 &   ---   \\
\hline
Fe & 0.06$\pm$ 0.01   & 0.08 $\pm$ 0.04  & 0.03 $\pm$ 0.01 & -2.4 $\pm$ 1.1  \\  
\hline
\end{tabular}
\end{center}
\caption{\small Parameters obtained in the fits of Eqs. \ref{mean}--\ref{sigma2} to simulation results.}
\label{tablemu}
\end{table}

The probability density of an event to belong to a  proton or iron $\eta$
distribution is given by the appropriate Gaussians:
\begin{equation} 
G=\frac{1}{\sqrt{2 \pi} \sigma_\eta(\theta_i,E_i)}
\exp(-\frac{(\eta_i-\mu_\eta(\theta_i,E_i))^2}{2 \sigma_\eta(\theta_i,E_i)^2})
\\
\label{prob}
\end{equation}
where $\theta_i$, $E_i$ and $\eta_i$ are the zenith angle, the energy and the
 reconstructed values of $\eta$ of each event. 

To fit the composition, assuming only two components as seems reasonable from
the discussion in \ref{method1}, we use the the maximum likelihood method. The quantity
to maximise in this method is:
\begin{equation}
\ln ~P(F_{\rm p})=\ln (P_1~P_2~....~P_n)=\sum_{i=1}^n~\ln~P_i \, ,
\label{likehood}
\end{equation}
where $P_i$ is given by:
\begin{equation}
P_i= F_{\rm p} G_{\rm p} + (1-F_{\rm p}) G_{\rm Fe}
\end{equation}
where $F_{\rm p}$ is the fraction of protons and $G_{\rm p}$ ($G_{\rm
Fe}$) is the probability of the event to be proton (iron), as given in
Eq. \ref{prob}.

\begin{figure}
\ybox{0.4}{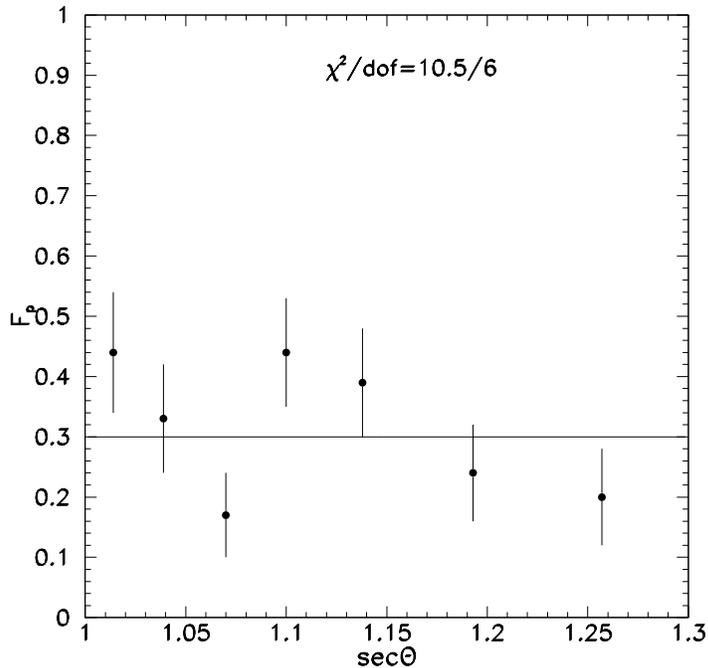}
\caption{\small Predicted value of $F_{\rm p}$ in the energy range
0.3--0.5 EeV for different zenith angle bins. The predicted value of
$F_{\rm p}$ does not depend on zenith angle.}
\label{zenfrac}
\end{figure}

With this procedure we can use events at all zenith angles to fit
$F_{\rm p}$ in a given energy range. As a demonstration of reliability
of the method, we have divided our set of data into bins of zenith
angle, selecting events between 0.3--0.5 EeV. For each angular bin we
fitted the composition using the method described
above. Fig. \ref{zenfrac} shows the results. As expected, the values
derived of $F_{\rm p}$ obtained do not depend on zenith angle. The
average value is in agreement with the value obtained in the previous
section (F$_{\rm p}=0.30 \pm 0.04$).
\begin{figure}
\ybox{0.4}{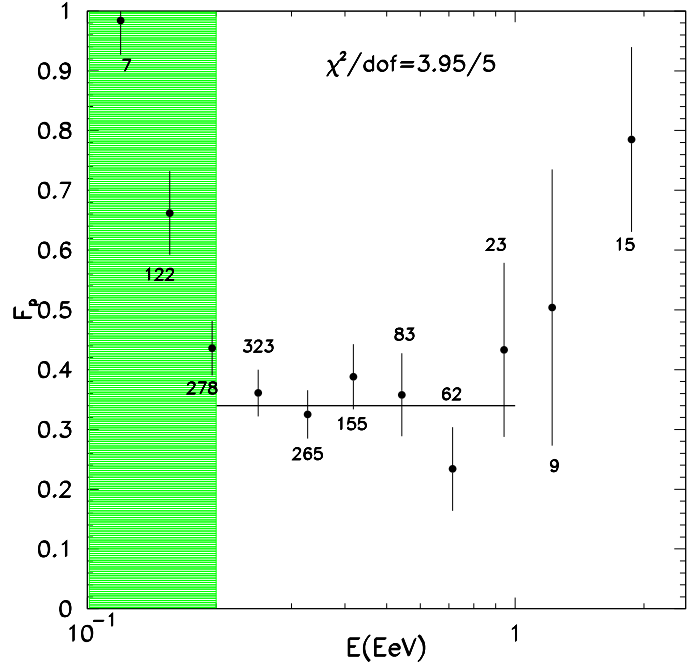}
\caption{\small Predicted value of $F_{\rm p}$ as a function of the
energy. A fit to a constant composition in the energy range 0.2--1.0 EeV
is also shown with its corresponding $\chi^2$. The number of events in
each energy bin is shown. The shadow region corresponds to the energy
range in which the analysis is affected by trigger biases (see text).}
\label{enfrac}
\end{figure}
\begin{figure}
\ybox{0.3}{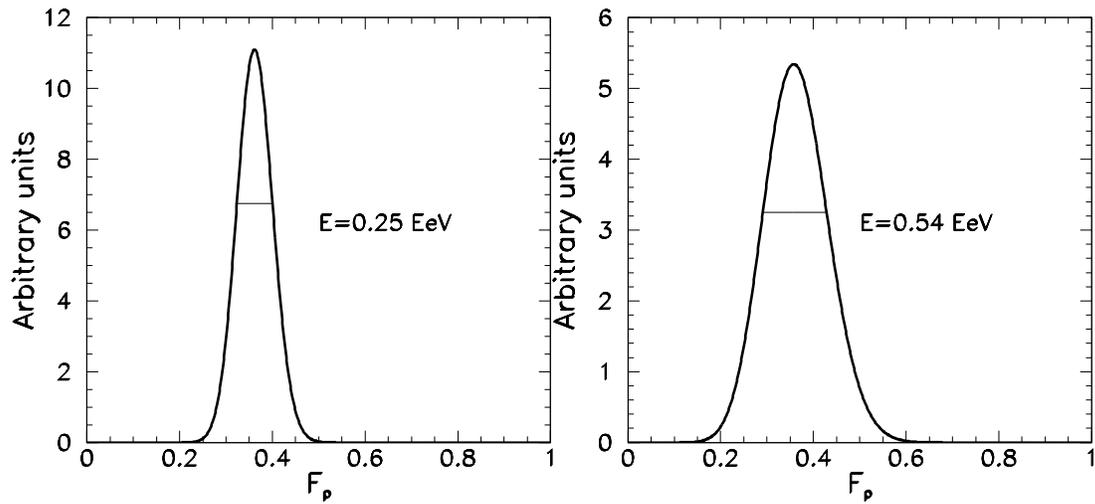}
\caption{\small Likelihood curves in two energy bands of Fig. \ref{enfrac}. The
horizontal lines correspond to a value of the maximum
of the log-likelihood function minus 0.5 (1 $\sigma$ level).}
\label{87_likelihood}
\end{figure}

We have also investigated the possibility of having oxygen and helium in
the cosmic ray beam. For this purpose we have selected events in the
energy range 0.3--0.5 EeV and with 1.1 $< \sec~\theta < 1.14$ , and,
using a modification of Eq. \ref{likehood}, fitted the fraction of
oxygen (helium) assuming a three component mixture of primaries: proton,
iron, and oxygen (helium). The fitted fraction of oxygen or helium are
less than 1\%. At 68\% confidence level the fraction of oxygen (helium)
should be less than 15\% (25\%).  Therefore, as in the previous section,
we will assume a dual component mixture of proton and iron for our final
discussion.

Fig. \ref{enfrac} shows the predicted fraction of protons in the cosmic
ray beam as a function of the energy. In this case we have used the
events from all zenith angles. It is apparent that in the energy range
0.2--1.0 EeV the composition does not change within the limits of
measurement. The rapid increase of $F_{\rm p}$ at energies below 0.2 EeV
is an artificial effect. It was established very early \cite{kioto} that
below 0.2 EeV steep showers have a higher probability to trigger the
array, and as the average proton shower is steeper than the average iron
shower, $F_{\rm p}$ increases rapidly at lower energies.  A fit to a
constant composition in the energy range 0.2--1.0 EeV gives a predicted
value of $F_{\rm p}=0.34 \pm 0.02$. Examples of the likelihood curves
used to extract mass composition are shown in Fig. \ref{87_likelihood}
for two energy bins of Fig. \ref{enfrac}.

We thus conclude that there is no evidence for any change of mass with
energy in the range studied and that Fe is the dominant component in
this energy range. The data of Fig. \ref{enfrac} do not exclude the
possibility of the proton component being enhanced at energies larger
than 1.0 EeV.

\subsection{Sensitivity to hadronic model}

\begin{figure}[b]
\ybox{0.4}{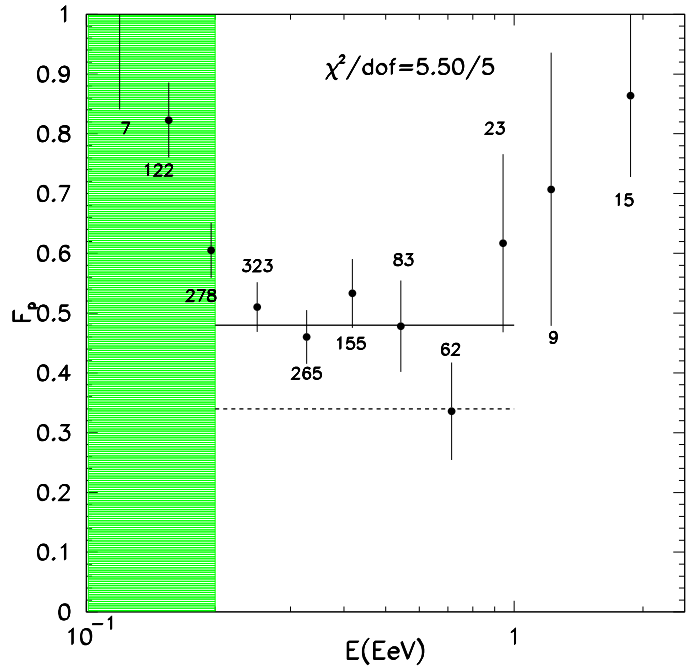}
\caption{\small Predicted value of $F_{\rm p}$ as a function of the energy for
QGSJET01 (solid line) and QGSJET98 (dashed line). The shadow region
corresponds to the energy range in which the analysis is affected by
trigger biases (see text).}
\label{enfrac2}
\end{figure}
The simulations presented in this work were obtained using CORSIKA and
QGSJET98 as high energy interaction model. Recently QGSJET98 has been
replaced by QGSJET01. The major change was a modification in the
treatment of diffractive processes \cite{heck}. This change reduced the
value of the position of the shower maximum $X_{\rm m}$ by $\approx$ 10
g/cm$^{2}$, that is showers develop higher in the atmosphere.

To establish the sensitivity of our results to this change in the
hadronic model we simulated 4 sets of 100 CORSIKA showers with the new
model for proton and iron primaries at a zenith angle of $26^\circ$ and
with primary energies of 0.4, and 0.8 EeV. We found that the spread of
$\eta$ ($\sigma_{\eta}$) does not change, but the value $\mu_{\eta}$ is
reduced by 0.04 for both proton and iron. Therefore, we can calculate
the predicted value of $F_{\rm p}$ as a function of the energy using the
likelihood method, but reducing the value of $\mu_\eta$ in
Eq. \ref{mean} by 0.04. Fig. \ref{enfrac2} shows the results compared
with the expectation obtained in the previous subsection. The value of
$F_{\rm p}$ is increased from 0.34 to 0.48.

The QGSJET01 model has been shown to have the smallest value of $X_{\rm
m}$ when compared to a variety of hadronic models ({\sc neXus} 2, SIBYLL
2.1, DPMJET II.5) \cite{heck}. Some of these models have the mean value
of $X_{\rm m}$ increased by $\approx$ 10 g/cm$^{2}$ compared to
QGSJET98. Assuming that the ratio of electromagnetic to muonic particles
in the shower for these models is the same than QGSJET98, we can
estimate the changes in our results in the manner used for the QGSJET01
model. The value $\mu_{\eta}$ is now increased by 0.04 and the value of
$F_{\rm p}$ is reduced from 0.34 (QGSJET98) to 0.23.

We can also estimate the changes in the risetime analysis when using
another hadronic model.  Using the relationship between X$_{\rm m}$ and
t$_{1/2}$ in \cite{walker}, a shift of $X_{\rm m}$ by $\approx$ 10
g/cm$^{2}$ corresponds to a change in t$_{1/2}$ at 400 m from the shower
core of less than 1 ns. Hence, the agreement between data and
simulations exemplified by Fig. \ref{risevsr} still holds.
 
As a final result, we state that the data on lateral distributions in
the energy range 0.2--1.0 EeV is well fitted by a dual p/Fe composition
with about (34 $\pm$ 2)\% of the signal being protons constant in this
energy range. The error in this quantity comes from the statistics of
the data available. The systematic uncertainty, introduced by the choice
of hadronic models, is $\approx$ 14\%, larger than the statistical
error. Clearly, further calculations, using a range of models, are
highly desirable.

\section{Comparison with other experiments and conclusions}

The Haverah Park data, in the light of new hadronic models have provided
a new estimation on the mass composition in the energy range 0.2--1.0
EeV. Efforts to understand the origin of cosmic rays at any energy are
greatly hampered by our lack of knowledge of the mass distribution in
the incoming cosmic ray beam. We have obtained here a prediction for the
mass composition in the energy range 0.2--1.0 EeV. We can compare only
with one experiment operating in the same energy range: the HiRes
prototype, operated with the MIA detector. Using the QGSJET98 model they
find a rapid change towards a light composition between 0.1--1.0 EeV
\cite{hires}, which is in contradiction with our results.  The question
of mass composition is thus far from being resolved and there is clearly
scope for alternative approaches.

Some years ago an analysis of the Fly's Eye data \cite{gaisser2} pointed
to a change from an iron dominated composition at $3 \times 10^{17}$ eV
to a proton-dominated composition near 10$^{19}$ eV.  This conclusion
was drawn from a study of the variation of depth of maximum with energy
(the elongation rate) and from an analysis of the spread in depth of
maximum at a given energy. The conclusions on mass composition from the
study of the spread in depth of maximum were confirmed by \cite{Wibig}
but none of these analyses extended to energies above 10$^{19}$ eV. A
different conclusion has been reached by the AGASA group based on the
variation of the muon content of showers with energy
\cite{qgsjet_uhecr}.  Their analysis favours a composition that remains
``mixed'' over the 10$^{18}$ to 10$^{19}$ eV decade. The statistics in
this energy range from this data set are too small to be able to claim
any agreement with either of the two experiments.  We are starting to
work on a risetime analysis in this energy range which could provide
some relevant information.

We can also compare our results with the mass composition obtained by
KASCADE. The mean value of the mass at their highest accessible energy
($\approx 10^{17}$ eV) was estimated to be $\langle \ln~A\rangle = 3.5
\pm 0.5$ \cite{kascade}, which should be compared with the result
obtained in this work which corresponds to $\langle\ln~A\rangle=2.65 \pm
0.08 \pm 0.6$ in the energy range 2--10 $\times 10^{17}$ eV. In contrast
to our results KASCADE also claims that protons have largely disappeared
at $10^{16}$ eV.

\subsection*{Acknowledgments:} 
Thanks are given to members of the Haverah Park group who helped to get
the data discussed here over 20 years ago.  In particular the major
contributions made by R.J.O. Reid, R.N. Coy and C.D. England are
gratefully acknowledged.  We are grateful to the staff of the Computing
Centre of IN2P3 in Lyon and to M. Risse for the help in producing the
simulated showers for this analysis.

\end{document}